\documentclass [twocolumn,aps,prb]{revtex4}
\usepackage{epsfig,graphicx,amssymb,color,amssymb,amsmath,bm}

\begin{document}
\title{Effect of periodic potential on exciton states in semiconductor carbon nanotubes}

\author{Oleksiy Roslyak$^1$}\email{oroslyak@fordham.edu}
\author{Andrei Piryatinski$^2$}\email{apiryat@lanl.gov}
\affiliation{$^1$Department of Physics and Engineering Physics, Fordham University, Bronx, NY 10458
\\$^2$Theoretical Division, Los Alamos National Laboratory, Los Alamos, New Mexico, 87545}

\date{\today}

\begin{abstract}

 We develop a theoretical background to treat exciton states of semiconductor single-walled carbon nanotubes (SWCNTs) in presence of a periodic potential induced by the surface acoustic wave (SAW) propagating along semiconducting SWCNT. The formalism naturally accounts for the electronic bands splitting into the Floquet sub-bands brought about by the Bragg scattering on the SAW potential.  Optically induced transitions within the Floquet states and formation of correlated electron-hole pairs, i.e., exciton states, are examined numerically. We discuss dynamical formation of new van Hove singularities within electron-hole continuum and associated reduction of the exciton oscillator strengths and its effect on the photoluminescence quenching in presence of the SAW.  We argue that SAW induced dynamical gaps in the single particle dispersion leads to redistribution of the oscillator strength from excitons to the Floquet edge states. The simulations also confirm exciton energy Stark red shift as well as reduction in the binding energy. Comparison of our results with previous theoretical and experimental studies is provided.  

\end{abstract}

\maketitle



\section{Introduction}
\label{Sec-Intro}

Single-walled carbon nanotubes (SWCNTs) are perfect quasi-1D structures demonstrating reach variety of transport and photophysical properties. \cite{Saito.Book:1998,Dresselhaus.Book:2001} They are promising candidates for building blocks in nanoelectronics, quantum information processing, opto-electronics, and light emitting devices. In contrast to semiconductor low-dimensional materials, SWCNTs electronic structure is described by the Dirac equation for right and left propagating quasi-spinors associated with the chiral Bloch states. The quantum interference and tunneling of these states can result in new effects that are not feasible in semiconductor nanostructures. \cite{CastroNeto:2009} Coulomb interactions in SWCNTs are substation making many-body electronic effects important in understand their photophysics. \cite{Dresselhaus:2007}

Common approach to control the transport and optical properties of semiconductor nanostructures, photonic materials and SWCNTs is based on modifications of their chemical composition and geometry. However, in many situations (e.g., tunable sensors and light emitting devices) it is desirable to control optical and transport properties dynamically by applying external electric and magnetic fields. A reliable way to produce and control the external potential  is to excite MHz through THz frequency surface acoustic waves (SAW) in a piezoelectric substrate holding the nanostructure.  Polarization of the substrate surface produces SAW-modulated electric field that couples to the materials electron gas modulating its wavefunction and as a result affecting their charge carriers transport and emission properties, particularly giving rise to the effect of sonoluminescence.\cite{Ostrovskii:1999}

Optical properties of various semiconductor nanostructures such as quantum dots, quantum wires and quantum wells influenced by SAW potential have been extensively examined.\cite{GarciaCristobal:2004,Kinzel:2011,Rocke:1997,Santos:1998,Volk:2010,Volk2011} In light of developing quantum photon sources, an ability to control single-photon emission by applying SAW has been demonstrated in nanowires.\cite{Hernandez:2011} In arrays of semiconductor quantum dots and rods, SAW results in the modulation of  the emission line and leads to the exciton charging that can be detected via sonoluminescence measurements.\cite{Gell:2008,Volk:2010,Volk2011} On the other hand,  use of SAW has been reported to control properties of photonic materials.\cite{Fuhrmann:2011,DeLima:2005} In semiconductors the SAW-induced fields form dynamically modulated band gap structures. In quantum wires and quantum wells this causes ptotoluminescence quenching via exciton dissociation into free electron and hole sates and their further transport to the emission centers resulting in spatial modulation of emission properties which is the essence of the sonoluminescence effect.\cite{Kinzel:2011,Rocke:1997,Santos:1998}  An attempt to describe the quenching by a conventional Stark effect predicts an essential excitonc red shift accompanying the reduction in exciton oscillator strength. However such shift has not been experimentally observed in core-shell nanowires\cite{Hernandez:2011}

The Dirac electronic structure of {\em metallic} SWCNTs  can also be influenced by an applied periodic  potential. Specifically,  such potential with a period much larger than SWCNT diameter causes formation of a Bragg grating that mixes the right and the left propagating Bloch states. This gives rise to the splitting of the valence and conduction band into Floquet sub-bands separated by dynamical small gaps.\cite{Talyanskii:2001,Novikov:2005} In light of developing new electronic devices including those for quantum information processing and metrology a problem of quantized transport \cite{Thouless:1983} can be addressed using SAW modulation of metallic SWCNTs. It has been demonstrated that the electric potential applied to contacts or impurity states in SWCNT can produce local potential walls forming the so-called quantum dots in which the charges get localized.  Application of SAW potential results in the Coulomb blockade assisted quantized charge pumping through the quantum dots.\cite{Buitelaar:2008,Leek:2005,Wurstle:2007}  

Recently, SAW induced photoluminescence quenching in a collection of {\it semiconductor}  SWCNTs has been examined experimentally. \cite{Regler:2013} From theory point of view the effect of {\it static} electric field on exciton properties in semiconductor SWCNTs has been considered.\cite{Perebeinos:2007}. However, such approach meets the same difficulties as in the case of nanowires, namely, the absence of a discernible red shift of exciton states.  This calls for development of a self-constant  theoretical approach treating the effect of periodic potential on non-interacting electron and hole states in semiconductor SWCNTs and subsequently on the correlated electron-hole pairs (i.e., the excites) and their optical properties. In Sec.~\ref{Sec-model}, we propose such a theoretical model. Subsequently, the model is used for numerical modeling of the  SAW effects on the lowest exciton states in semiconductor-SWCNTs  whose result we discuss in Sec.~\ref{Sec-Res}.  Conclusions are drawn in Sec.~\ref{Sec-Conc}.

	
\section{Model for exciton states modulated by periodic potential}
\label{Sec-model}

In our calculations we apply band folding approach to a semiconductor infinitely long SWCNT oriented along $y$-direction. Periodical boundary conditions are applied around the circumference direction defined by the $x$-axis of the lab coordinate system with the period of $L_x$. The SAW is launched by a transducer on the piezoelectric surface along the tube as described in Ref.\cite{Hernandez:2011}. In the moving frame of reference the pezoelectric potential along the tube can be expressed as:  
\begin{equation}
\label{SAW-potential}
U (y)= U_0 \cos(G_0 y),
\end{equation}
 where the SAW wave vector is $G_0=2\pi/L_y$, and $L_y$ is the SAW period. The length of SWCNT is set to be multiple of $L_y$. The effect of SAW transverse component is neglected.

In our calculations of bound exciton states, we follow the formalisms originally proposed by Ando.\cite{Ando:1997,Ando:2005} This formalism requires: electronic band structure (dynamic or otherwise), screened  electron-electron Coulomb interactions as well as their correlations. those ingredients are combined into a Bethe-Salpeter (BS) equation in reciprocal space governing bound exciton states well separated form uncorrelated quasiparticle continuum in semiconducting SWCNTs. Below we follow these steps and blend in the SAW potential \eqref{SAW-potential} into the BS equation. Once the wavefunction and eigenenergies for the quasiparticle states are know the basic optical properties can be evaluated based on the dynamical conductivity.  

\subsection{Non-interacting electron states. Floquet super-lattice.}

In the vicinity of $K$ and $K'$  points associated with graphene Brillouin zone, SWCNT non-interacting electron weave function is sought in the form
\begin{equation}
\label{eWF-ansatz}
{\bm\Psi}(x,y) =\bm\Phi(y)\frac{1}{\sqrt{L_x}} \texttt{e}^{-i k_{x}(n,\nu) x},
\end{equation}
where 
\begin{equation}
\label{kx}
k_x(n,\nu) =\frac{2\pi}{L_x}\left(n-\frac{\nu}{3}\right),
\end{equation}
is quantized circumference wave vector depending on integer angular momentum quantum number $n$ and the valley index $\nu=\pm 1$ associated with the Brillouin zone $K$ ($K'$) point.\footnote{The case when $\nu=0$ corresponds to metallic SWCNT and is not considered below.} The envelope function, $\bm\Phi(y)$, is a two-component spinor satisfying the 1D Dirac equation, gauge transformed into \cite{Novikov:2005,CastroNeto:2009}:
\begin{widetext}
\begin{eqnarray}
\label{enfEQ}
\left({
\begin{matrix}
-i \hbar \text v_F\partial_y - E & -i \hbar \text v_F k_x(n,\nu) \texttt{e}^{i \phi(y)}\\
 i \hbar \text v_F k_x(n,\nu) \texttt{e}^{-i \phi(y)} & i \hbar \text v_F\partial_y -E
\end{matrix}
}\right)
\boldsymbol{\Phi}(y)=0,
\end{eqnarray}
\end{widetext}
with $\hbar \text v_F = 6.46$~eV$\AA$ and $\text v_F$ being the Fermi velocity. The SAW periodic potential in the above equation enters via the phase modulation, $e^{i\phi(y)}$, of the kinetic term in Eq.~(\ref{enfEQ}) with the phase given by:
\begin{equation}
\phi (y)=\frac{2}{\hbar \text v_F} \int \limits_0^y d\xi U(\xi)=u_0\int \limits_0^y d\xi \cos\xi,
\end{equation}
, which linearly depends on the dimensionless SAW potential amplitude 
\begin{equation}
\label{Uamp}
u_0=\frac{2U_0}{\hbar \text v_F G_0}.
\end{equation}

To account for this phase effect, we expend the exponent in terms of the first kind Bessel functions, $e^{i\phi(y)}=\sum \limits_{l=-\infty}^\infty J_l (u_o)e^{i l G_0 y}$, and  further seek the envelope function in the form of Floquet series\cite{Gu:2011,Perez:2014}
\begin{equation}
\label{envFr}
\boldsymbol{\Phi}^\nu_{k}(y) =
\frac{1}{\sqrt{L_y}}
\sum \limits_{l=-\infty}^\infty
\left({
\begin{matrix}
A_{k-l G_0}\\
B_{k-l G_0}
\end{matrix} 
}\right)
e^{-i(k-l G_0)y}.
\end{equation} 
This results in the following set of eigenvalue equations for the expansion coefficients  
\begin{widetext}
\begin{eqnarray}
\label{eqA}
 \left({ -\hbar \text v_F(k - l G_0)-E}\right)A_{k - l G_0} 
	-i \hbar \text v_F k_{x}(n,\nu) \sum \limits_{l'=-\infty}^\infty J_{l'}(u_0)  B_{k - (l-l')G_0}&=&0,
\\\label{eqB}
 \hbar \text v_F k_{x}(n,\nu) \sum \limits_{l'=-\infty}^\infty J_{l'}(u_0) A_{k - (l+l')G_0}
	+\left({ \hbar \text v_F(k - l G_0)-E}\right)B_{k - l G_0}&=&0.
\end{eqnarray}
\end{widetext}
 
 Eqs.~(\ref{eqA}) and (\ref{eqB}) can be solved numerically, resulting in quantized eigenenergies, $E^\nu_{s,n,k}$ and eigenvector components 
\begin{equation}
\label{single-F}
\bm F^\nu_{s,n,k-lG_0}=
\left({
\begin{matrix}
A^{\nu}_{s,n,k-lG_0}\\
B^{\nu}_{s,n,k-l G_0}
\end{matrix} 
}\right).
\end{equation}
satisfying the orthonormality  relationship $\sum_{l=-\infty}^\infty ({\bm F^\nu}^*_{s,n,k-lG_0}\cdot\bm F^\nu_{s',n,k-lG_0})=\delta_{s,s'}$. Here, we introduce a new integer quantum number $s$ accounting for the Bragg scattering at the ends of SAW induced Brillouin sub-zone, $0 \le k\le lG_0$. This quantization results in the splitting of the valence and conduction bands into the so-called Floquet sub-bands (also known as replicas). Specifically, $s$ acquiring even (odd) values numerate the Floquet sub-bands within the valence (conduction) band. Finally, spatial distribution of the single particle wavefunctions (Eq.~(\ref{eWF-ansatz})) can be written in a short form (implicit summation on $l$ index is assumed)  as
\begin{widetext}
\begin{equation}
\label{single-wf}
\bm\Psi^\nu_{s,n,k-lG_0}(x,y) =\frac{1}{\sqrt{L_x L_y}} 
\bm F_{\alpha,k_x,k-lG_0} e^{-i \frac{2\pi}{L_x}(n-\frac{\nu}{3}) x-i(k-l G_0)y}.
\end{equation}
\end{widetext}
 
\subsection{Screened Coulomb interaction}

To account for Coulomb screening effect, we introduce a real space nonlocal dielectric function defined by the following integral expression 
\begin{equation}
\label{eps-eq}
\epsilon \left(\bm r_1,\bm r_2\right) = \delta\left(\bm r_1-\bm r_2\right) 
-\int d^2 \bm r V\left(\bm r\right) \Pi \left(\bm r_1-\bm r,\bm r_2\right),
\end{equation}
where the coordinate vector is $\bm r=(x,y)$ and two-dimensional bare Coulomb interaction operator is~\cite{Ando:1997} 
\begin{widetext}
\begin{equation}
\label{Vee-def}
V\left(\bm r\right)=
\frac{2\pi e^2}{\epsilon_\infty  L_x L_y} \sum_q e^{i q y}
K_0 \left(  \frac{L_x|q|}{\pi} \left\vert\sin\left(\frac{x}{4\pi L_x}\right)\right\vert\right).
\end{equation}
\end{widetext}
The operator depends on the electron charge, $e$, static part of the environment's dielectric function at optical frequency, $\epsilon_\infty$, and modified Bessel function of the second kind, $K_0$. 

The nonlocal polarization operator, $ \Pi \left(\bm r,\bm r'\right)$, entering Eq.~(\ref{eps-eq}) can be represented as~\cite{Kushwaha:2001}
\begin{equation}
\label{Plz-exp}
\Pi \left(\bm r_1,\bm r_2\right) = \sum_{\alpha_1,\alpha_2}\Pi^0_{\alpha_1\alpha_2}
\rho_{\alpha_1\alpha_2}(\bm r_1)\rho_{\alpha_1\alpha_2}^*(\bm r_2),
\end{equation}
where the resonance component is 
\begin{equation}
\label{Plz-0}
\Pi^0_{\alpha_1\alpha_2} = 2 \frac{f\left({E_{\alpha_1}}\right) - f\left({E_{\alpha_2}}\right)}{E_{\alpha_1} - E_{\alpha_2}}
g\left({E_{\alpha_1}}\right)  g\left({E_{\alpha_2}}\right) 
\end{equation}
and contains the eigenenergies $E_{\alpha_i}$, Fermi-Dirac distribution function, $f(E)$ with Fermi energy set to zero, a smooth cutoff function, $g(E)$, selecting contribution only from the states that belong to the vicinity of the Fermi level, and the prefactor two reflecting spin degeneracy. In our particular case of interest, the electron density matrix elements, entering Eq.~(\ref{Plz-exp}) can be expressed in terms of non-interacting electron wave function introduced in Eqs.~(\ref{single-F}) and (\ref{single-wf}). Specifically for fixed $K$ or $K'$ points, one can write down   
\begin{equation}
\label{rho_r}
\rho^\mu_{\alpha_1\alpha_2}(\bm r) = \left({\bm\Psi^\nu}^*_{\alpha_1}(\bm r) \cdot \bm\Psi_{\alpha_2}(\bm r)\right),
\end{equation}
where composite index contains whole set of quantum numbers $\alpha_i=\{s_i,n_i,k_i-l_iG_0\}$ except $\nu$ that is fixed. The substitution of Eqs.~(\ref{Plz-0}) and (\ref{rho_r}) along with Eq.~(\ref{single-wf}) into Eq.~(\ref{Plz-exp}) results in the following representation of the polarization operator in the coordinate space
\begin{widetext}
\begin{eqnarray}
\label{Plz-crd}
\Pi \left(\bm r_1-\bm r_2\right) = \frac{1}{L_xL_y}\sum_{\alpha_1,\alpha_2}\Pi^\nu_{\alpha_1\alpha_2}
e^{i\frac{2\pi}{L_x}(n_1-n_2)(x_1-x_2)}
e^{i(k_1-l_1G_0-k_2+l_2G_0)(y_1-y_2)}
\end{eqnarray}
\end{widetext}
where the operator matrix element reads 
\begin{equation}
\label{Plz-mtrx}
\Pi^\nu_{\alpha_1\alpha_2} = 2 \frac{f\left({E_{\alpha_1}}\right) - f\left({E_{\alpha_2}}\right)}{E_{\alpha_1} - E_{\alpha_2}}
g\left({E_{\alpha_1}}\right)  g\left({E_{\alpha_2}}\right) 
\left|{\bm F^\nu_{\alpha_1}}^*\cdot \bm F^\nu_{\alpha_2}\right|^2.
\end{equation}

Taking advantage of the polarization operator (Eq.~(\ref{Plz-crd})) dependens on the coordinate difference, one can substitute it into Eq.~(\ref{eps-eq}) and perform integration over $d\bm r$, resulting in the expression for the non-local dielectric function that depends on the coordinate difference, i.e., $\epsilon(\bm r_1-\bm r_2)$. Further introducing $(x,y)=\bm r=\bm r_1-\bm r_2$ and the following Fourier transform 
\begin{widetext}
\begin{equation}
\label{eps-Four-int}
\epsilon_{n_1-n_2}(q)=\int_{-L_x/2}^{L_x/2}\frac{dx}{L_x}\int_{-L_y/2}^{L_y/2}\frac{dy}{L_y}
\epsilon(x,y)~e^{-i\frac{2\pi}{L_x}(n_1-n_2)x-iqy}
\end{equation}
\end{widetext}
one can find the Fourier transform of the r.h.s. of Eq.~(\ref{eps-eq}) with polarization operator given by Eq.~(\ref{Plz-crd}). This results in the following equation
\begin{widetext}
\begin{equation}
\label{epsq-eq}
\epsilon^\nu_{n_1-n_2}(q)=1+ \frac{2 e^2}{\epsilon_\infty} I_{\vert{n\alpha-n\beta}\vert}\left({\frac{L_x\vert{q}\vert}{2 \pi}}\right) K_{\vert{n\alpha-n\beta}\vert}\left({\frac{L_x\vert{q}\vert}{2 \pi}}\right)\Pi^\nu_{n_1-n_2}(q)
\end{equation}
\end{widetext}
where the Fourier transform of the polarization operator reads
\begin{eqnarray}
\label{Plz-q}
\Pi^\nu_{n_1-n_2}(q)&=&\sum_{\alpha_1'\alpha_2'}\delta_{n_1-n_2;n_1'-n_2'}
\\\nonumber &\times&
\delta_{k_1'+q-l_1'G0;k_2'-l_2'G_0}
\Pi^\nu_{\alpha_1'\alpha_2'}
\end{eqnarray}
with $\Pi^\nu_{\alpha_1'\alpha_2'}$ given by Eq.~(\ref{Plz-mtrx}).

According to the RPA formalism, screened Coulomb interaction in real space can be evaluated as
\begin{equation}
\label{VC-conv}
W\left(\bm r_1-\bm r_2\right)=
\int d^2 \bm r~\epsilon^{-1}\left({\bm r_1-\bm r_2-r}\right) V\left({\bm r}\right)
\end{equation}
where bare Coulomb operator, $V(\bm r)$, is given by Eq.~(\ref{Vee-def}) and the inverse dielectric function can be represented as
\begin{widetext}
\begin{equation}
\label{eps-Four}
\epsilon^{-1}(\bm r_1-\bm r_2)=\sum_{n_1,n_2,q} \frac{1}{\epsilon^\nu_{n_1-n_2}(q)}
e^{-i\frac{2\pi}{L_x}(n_1-n_2)(x_1-x_2)-iq(y_1-y_2)}.
\end{equation}
\end{widetext}
Here, $\epsilon^\nu_{n_1-n_2}(q)$ is determined by Eqs.~(\ref{epsq-eq}) and (\ref{Plz-q}). The substitution of Eq.~(\ref{eps-Four}) into Eq.~(\ref{VC-conv}) and subsequent evaluation of the integral over $d\bm r$ results in the following coordinate space expression for the screened Coulomb interaction
\begin{widetext}
\begin{eqnarray}
\label{W-srd}
W\left(\bm r_1-\bm r_2\right)=\sum_{n_1,n_2,q}\frac{2 e^2}{\epsilon_\infty\epsilon^\nu_{n_1-n_2}(q)} I_{\vert{n_1-n2}\vert}\left({\frac{L_x\vert{q}\vert}{2 \pi}}\right) 
K_{\vert{n_1-n2}\vert}\left({\frac{L_x\vert{q}\vert}{2 \pi}}\right)
e^{-i\frac{2\pi}{L_x}(n_1-n_2)(x_1-x_2)-iq(y_1-y_2)}.
\end{eqnarray}
\end{widetext}

Finally, defining screened Coulomb interaction matrix elements as
\begin{eqnarray}
\label{W-mtrx}
W_{\alpha_1\alpha_2;\alpha_3\alpha_3}&=&\int d\bm r_1\int d\bm r_2 W\left(\bm r_1-\bm r_2\right)
\\\nonumber &\times&
\rho_{\alpha_1\alpha_3}(\bm r_1)\rho_{\alpha_2\alpha_4}(\bm r_2)
\end{eqnarray}
where the non-interacting electron densities are given by Eq.~(\ref{rho_r}). Using explicit form for the wave function in Eq.~(\ref{single-wf}) and subsequently evaluating the integrals over $d\bm r_1$ and $d\bm r_2$ and dropping summation over $q$, we arrive at the following expression for the screened Coulomb matrix element 
\begin{eqnarray}
\label{W-q}
&~&W^\nu_{\alpha_1\alpha_2;\alpha_3\alpha_3}(q)=\delta_{n_1-n_2;n_3-n_4}
\\\nonumber&\times&
\delta_{k_1+q-l_1G0;k_3-l_3G_0}\delta_{k_2-q-l_2G0;k_4-l_4G_0}
\\\nonumber&\times&
\frac{2 e^2}{\epsilon_\infty\epsilon^\nu_{n_1-n_2}(q)}I_{\vert{n_1-n_2}\vert}\left({\frac{L_x\vert{q}\vert}{2 \pi}}\right) K_{\vert{n_1-n_2}\vert}\left({\frac{L_x\vert{q}\vert}{2 \pi}}\right)
\\\nonumber&\times&
\left({\bm F^\nu_{\alpha_1}}^*\cdot \bm F^\nu_{\alpha_2}\right)\left(\bm F^\nu_{\alpha_3}\cdot {\bm F^\nu_{\alpha_4}}^*\right)
\end{eqnarray}
that can be numerically evaluated with the help of Eqs.~(\ref{epsq-eq}) and (\ref{Plz-q}). The matrix elements of bare Coulomb interaction have simple relation to the screened ones, $V^\nu_{\alpha_1\alpha_2;\alpha_3\alpha_3}(q)=\epsilon^\nu_{n_1-n_2}(q)W^\nu_{\alpha_1\alpha_2;\alpha_3\alpha_3}(q)$  ,  reading
\begin{eqnarray}
\label{V-q}
&~&V^\nu_{\alpha_1\alpha_2;\alpha_3\alpha_3}(q)=\delta_{n_1-n_2;n_3-n_4}
\\\nonumber&\times&
\delta_{k_1+q-l_1G0;k_3-l_3G_0}\delta_{k_2-q-l_2G0;k_4-l_4G_0}
\\\nonumber&\times&
\frac{2 e^2}{\epsilon_\infty}I_{\vert{n_1-n_2}\vert}\left({\frac{L_x\vert{q}\vert}{2 \pi}}\right) K_{\vert{n_1-n_2}\vert}\left({\frac{L_x\vert{q}\vert}{2 \pi}}\right)
\\\nonumber&\times&
\left({\bm F^\nu_{\alpha_1}}^*\cdot \bm F^\nu_{\alpha_2}\right)\left(\bm F^\nu_{\alpha_3}\cdot {\bm F^\nu_{\alpha_4}}^*\right).
\end{eqnarray}


\subsection{BS equation for Floquet quasiparticles. Bound exciton states}

Excitons are defined as bound quasiparticles separated from the electron-hole continuum by the binding energy resulted from the Coulomb direct and exchange interactions.
To obtain their eigenstate equation we adopt second quantization representation.  The real space field operator associated with the non-interacting  electron state wave functions defined (Eqs.~(\ref{single-F}) and (\ref{single-wf})) for a fixed $K$ or $K'$ point ($\nu=1$ or $\nu=-1$, respectively)  reads
\begin{equation}
\label{Psi-fop}
\hat\Psi^\nu (\bm r) = \sum_{\alpha} \Psi^\nu_\alpha (\bm r) \hat  c^\nu_{\alpha} ,
\end{equation}
 Accordingly, the creation, $c^{\nu\dag}_{\alpha}$  and annihilation,  $c^\nu_{\alpha}$, operators of states $|\alpha,\nu\rangle$ satisfy the Fermi commutation rules. Let us remind that the summation runs over all associated quantum numbers, denoted for brevity by the index $\alpha=\{s,n,k-lG_0\}$, except already fixed $\nu$.  The Hamiltonian for the interacting valence and conduction band electronic states can be obtained via straightforward  integration of a kinetic (quadratic in field operators) and Coulomb  interaction (Eq.~(\ref{Vee-def})) quartic in field operators terms. This results in 
\begin{eqnarray}
\hat H^\nu &=&
\sum_{\alpha} E^\nu_\alpha \hat c_\alpha^{\nu\dag} \hat c^\nu_\alpha 
\\\nonumber
&+& \frac{1}{2} \sum \limits_{\alpha_1\alpha_2;\alpha_3,\alpha_4}\sum_q V^\nu_{\alpha_1\alpha_2;\alpha_3,\alpha_4}(q) 
	\hat c^{\nu \dag}_{\alpha_1} \hat c^{\nu \dag}_{\alpha_2}\hat c^\nu_{\alpha_3} \hat c^\nu_{\alpha_4},
\end{eqnarray}
where $E_\alpha^\nu$ and $ V^\nu_{\alpha_1\alpha_2;\alpha_3,\alpha_4}(q) $ denote the eigenvalues of Eq. ~(\ref{eqA}) and bare Coulomb interaction matrix elements  given by Eq~(\ref{V-q}), respectively. 

Let us define the ground state of SWCNT associated with $K$ or $K'$ points as filled valance band by non-interacting electrons, i.e., the Hartree-Fock state,
\begin{equation}
\label{gX}
|g^\nu \rangle = \prod \limits_{\alpha_{\texttt v}} c^{\nu\dag}_{\alpha_{\texttt v}}|0\rangle ,
\end{equation}
where $|0\rangle$ denotes the vacuum, and index $\alpha_{\texttt v}$ specifies that the product is taken over all valance band states. The later are selected by remaining only those  $\alpha=\{s,n,k-lG_0\}$ whose index $s$ is odd integer. Further we introduce exciton creation operator describing  promotion of a $K$  ($K'$) point valence band electron to a $K$  ($K'$) point state in the conduction band as
\begin{equation}
\label{BX}
\hat{B}^\dag_{\lambda} = \sum \limits_{\alpha_{\texttt c},\alpha_{\texttt v}} 
\Phi^\lambda_{\alpha_{\texttt c},\alpha_{\texttt v}} \hat{c}^\dag_{\alpha_{\texttt c}}\hat{c}_{\alpha_{\texttt v}},
\end{equation}
and associated exciton state as 
\begin{equation}
\label{eigenX}
|\lambda\rangle  = \hat{B}^\dag_{\lambda}|g\rangle.
\end{equation}
Notice that we dropped index $\nu$ for the excitons states. This reflects the fact that the excitons are formed from the electron and hole states that both belong to the same $K$ or $K'$ point and have the center of mass momentum zero, i.e., form $\Gamma$-point states. These states are two-fold degenerate reflecting that $K$ and $K'$ point excitons are indistinguishable.\cite{Dresselhaus:2007}   

Heisenberg equation of  notion for the exciton operators oscillating with corresponding eigenfrequencies leads to the following equation for the exciton states
\begin{equation}
\label{Heis-eq}
\left[{\hat H^\nu,\hat{B}^\dag_{\lambda}}\right]\vert{g}\rangle = E_{\lambda}\hat{B}^\dag_{\lambda}\vert{g}\rangle.
\end{equation}
Multiplying both sides of Eq.~(\ref{Heis-eq}) by the ket $\langle g | \hat c^{\nu\dag}_{\alpha_{\texttt v}} \hat c^\nu_{\alpha_{\texttt c}}$ and subsequently performing the normal ordering of the Fermion operators, one obtains the following Bether-Salpeter equation for the exciton (i.e., correlated electron-hole pair) eigenfunctions, $\Phi^\lambda_{\alpha_{\texttt c},\alpha_{\texttt v}}$, and energies, $E_\lambda$,

\begin{widetext}
\begin{eqnarray}
\label{X-HF}
\sum_{\alpha_{\texttt v}',\alpha_{\texttt c}'} \left[ 
\left( \tilde{E}^\nu_{\alpha_{\texttt c}}- \tilde{E}^\nu_{\alpha_{\texttt v}} \right) 
\delta_{\alpha_{\texttt c},\alpha_{\texttt c}'} \delta_{\alpha_{\texttt v},\alpha_{\texttt v}'}
-\sum_q\left( W^\nu_{\alpha_{\texttt c}\alpha_{\texttt c}',\alpha_{\texttt v}\alpha_{\texttt v}'}(q)
- 2V^\nu_{\alpha_{\texttt c}\alpha_{\texttt v},\alpha_{\texttt c}'\alpha_{\texttt v}'}(q)\right) \right]
\Phi^\lambda_{\alpha_{\texttt c}'\alpha_{\texttt v}'} 
= E_\lambda \Phi^\lambda_{\alpha_{\texttt c}\alpha_{\texttt v}},
\end{eqnarray} 
\end{widetext}
where the renormalized single-electron energy reads 
\begin{equation}
\label{E-HF}
\tilde{E}^\nu_{\alpha_{i}} = E_{\alpha_{i}}
-\sum_{\alpha_{ i}'} \sum_q
V^\nu_{ \alpha_{ i} \alpha_{ i}'; \alpha_{ i} \alpha_{i}'} (q)  f(E_{\alpha_{i}'}),
\hspace{0.5cm} i={\texttt c}, {\texttt v};
\end{equation} 
and the screened, $W^\nu_{\alpha_{\texttt c}\alpha_{\texttt c}',\alpha_{\texttt v}\alpha_{\texttt v}'}$, and the bare, $V^\nu_{\alpha_{\texttt c}\alpha_{\texttt c}',\alpha_{\texttt v}\alpha_{\texttt v}'}$, Coulomb interaction matrix elements  are given by Eqs.~(\ref{V-q}) and (\ref{W-q}), respectively. In Eq.~(\ref{E-HF}), $f(E)$ denotes the Fermi-Dirac function. 

Equations~(\ref{X-HF}) and (\ref{E-HF}) have the structure of Bethe-Salpeter equation proposed to treat the exciton states in SWCNTs.\cite{Ando:1997,Ando:2005,Dresselhaus:2007,Jiang:2007,Spataru:2004} In the derivation above, we accounted for the effect of periodic SAW-induced potential, which modifies the structure of the wave functions (Eqs.~(\ref{single-F}) and (\ref{single-wf})) and subsequently the Coulomb matrix elements (Eqs.~(\ref{V-q}) and (\ref{W-q})) entering Eqs.~(\ref{X-HF}) and (\ref{E-HF}). In the limit of no periodic potential, $u_0=0$, and $G_0=a_{CC}$ (where $a_{CC}$ is the C-C bond length) only the terms with $l=0$ survive in Eqs.~(\ref{eqA}) and (\ref{eqB}) indicating that the splitting into sub-bands vanishes. This restricts the eigenfunction vector, Eq.~(\ref{single-F}), to two-component spinor and the quantum number $s=1,2$, exactly reproducing the  non-interacting wave-function first obtained by Ando.\cite{Ando:1997} Further use of two-component wavefunction for the evaluation of Coulomb interaction terms recovers the well-know Bethe-Salpeter used in the literature. In our numerical calculations discussed in Sec.~\ref{Sec-Res}, we used the limit of $u_0=0$ to check our numerical methodology.

\subsection{Optical observables}

Neglecting the depolarization effects the dynamical conductivity (current-current response) is \cite{Ando:1997,Ando:2005}
\begin{equation}
\label{sigma_yy}
\sigma_{yy}(E) = \frac{2 \hbar}{L_x L_y} \sum_{\nu=\pm 1} \sum \limits_{\lambda} 
\frac{-2 i E \left|\langle\lambda| \hat j^\nu_y  |g \rangle\right|^2}{E_{\lambda}
\left[{E^2_{\lambda} -E^2 - 2 i E \gamma}\right]}
\end{equation}
where $E_\lambda$ denotes the eigenenergies of the Bethe-Salpeter equation (Eqs.~(\ref{X-HF}) and (\ref{E-HF})) and $\gamma$ is associated dephasing rate, the prefactor of two reflects spin degeneracy. The current operator along SWCNT direction can be defined in terms of electron charge, $e$, Fermi velocity, $\text v_F$, the Pauli matrix, $\hat\sigma_y$,  and the  field operators for non-interacting electrons given by Eq.~(\ref{Psi-fop}) as
\begin{equation}
\label{jy-op}
\hat j^\nu_y = e \text v_F\int d \bm r \hat \Psi^{\nu\dag} (\bm r) \hat\sigma_y \hat\Psi^\nu(\bm r) 
\end{equation}
Direct calculation of the integral over $d\bm r$, and further use of Eqs.~(\ref{gX})--(\ref{eigenX}) results in the following representation of the exciton current matrix element entering Eq.~(\ref{sigma_yy})  
\begin{equation}
\label{jX}
\langle\lambda| \hat j^\nu_y  |g \rangle= \sum_{\alpha_{\texttt c}\alpha_{\texttt v}}\delta_{n_{\texttt c},n_{\texttt v}}
j^\nu_{\alpha_{\texttt c}\alpha_{\texttt v}}\Phi_{\alpha_{\texttt c}\alpha_{\texttt v}}^{\lambda *}.
\end{equation}
with the exciton eigenfunction defined by Eq.~(\ref{X-HF}) and the non-interacting electron transision current matrix element
\begin{equation}
\label{je-mtrx}
j^\nu_{\alpha_{\texttt c}\alpha_{\texttt v}}= e\text v_F\delta_{n_{\texttt c},n_{\texttt v}}
\delta_{k_{\texttt c}-l_{\texttt c}G_0,k_{\texttt v}-l_{\texttt v}G_0}
(\bm F_{\alpha_{\texttt c}}^{\nu *}\cdot\hat\sigma_y\bm F_{\alpha_{\texttt v}}^\nu)
\end{equation}
depending on the wave functions given by Eq.~(\ref{single-F}).

Finally, the absorption and emission spectra can be evaluated as 
\begin{equation}
\label{Spct}
S(E) \sim {\text Re}\{\sigma_{yy}(E)\},
\end{equation}
in which the oscillator strength of the optical transition can be found according to the following expression
\begin{equation}
\label{fX}
f_{\lambda} = \frac{2\hbar^2|\langle \lambda| \hat j^\nu_y  |g \rangle|^2}{ e^2 E_{\lambda} \gamma L_x L_y}.
\end{equation}

\begin{figure}[t]
  \centering
  \includegraphics[width=0.8\columnwidth,clip]{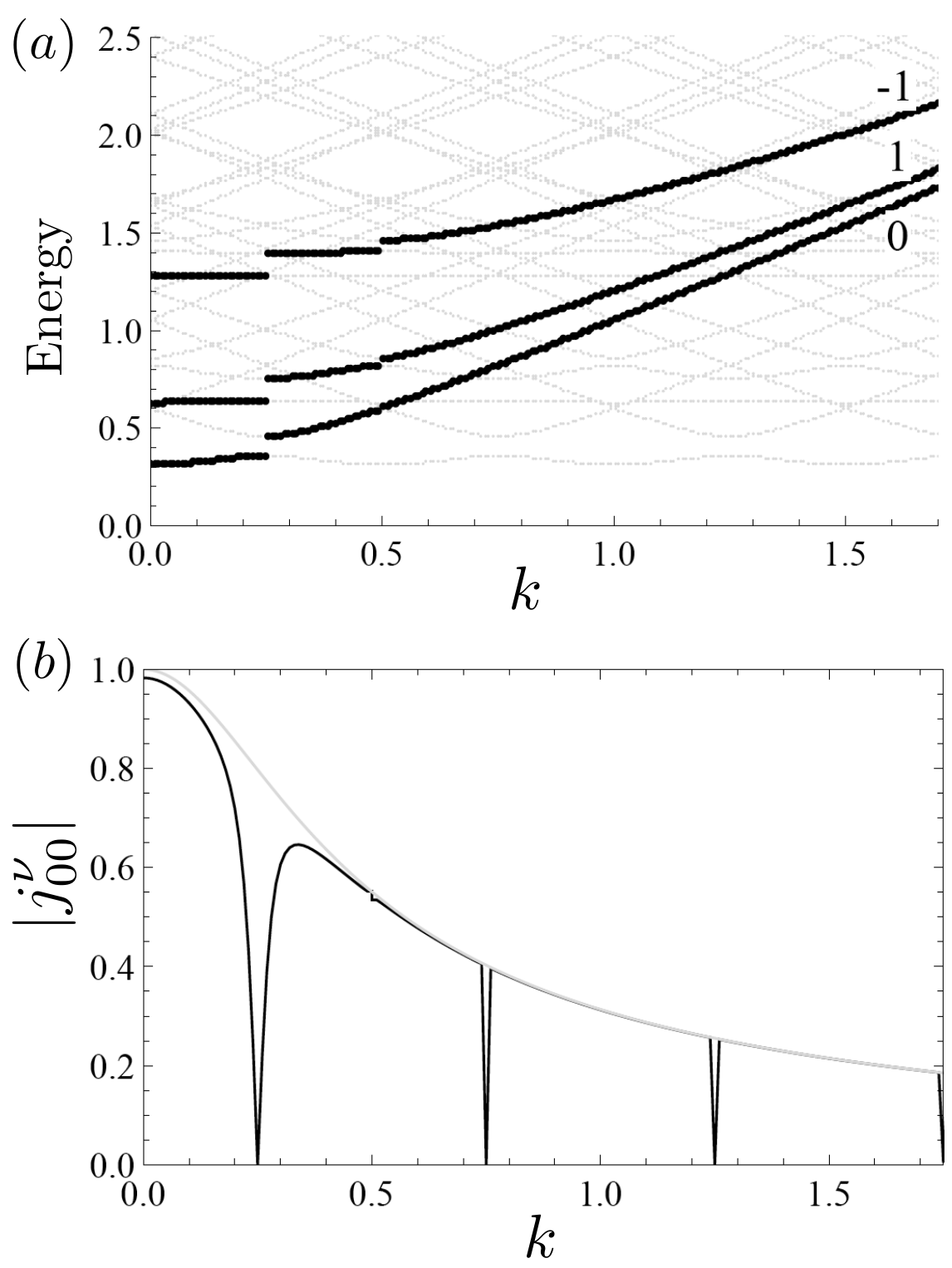}
  \caption{
(a) Non-interacting conduction band electron state dispersion in the presence of SAW induced potential with the period $L_y=10 L_x$. The energy is plotted in units of  $\hbar \text v_F 2 \pi/L_x$.   Grey curves represent Floquet sub-band structure due to the electrons Bragg scattering. Black curve describes the Floquet states originating from SAW unperturbed ($u_0=0$) SWCNT conduction bad electron states characterized by the quantum number $n_c=-1,0,1$. The spectrum is the same for $\nu=\pm 1$. The valance band spectrum has the same structure except the energies should be replaced with their negative values adjusted to proper band gap offset. Optical transitions are allowed between the valance and conduction band Floquet states marked by black lines with $n_{\texttt c}=n_{\texttt v}$.  
(b) Dispersion of the transition current matrix element $|j^\nu_{0,0}|$ defined in Eq.~(\ref{j00}) between valance and conduction band Floquet states marked by solid black curve with  $n_{\texttt c}=n_{\texttt v}=0$. Black (grey) curve shows the transition current values in the presence (absence) of SAW potential with the SAW period $L_y=100L_x$. In both panels the wave vector $k$ is given in units of $2\pi/L_x$ and the dimensionless SAW induced potential amplitude is set to $u_0=0.2$.
}
  \label{Fig:1}
\end{figure}

\section{Numerical results and discussion}
\label{Sec-Res}

We start by examining the structure of non-interacting electron conduction and valence band states obtained via numerical solution of Eqs.~(\ref{eqA}) and (\ref{eqB}). The calculated conduction band structure in the presence of SAW potential is show in Fig.~\ref{Fig:1}a. The valence band electrons have the same dispersion relationship but negative energies adjusted to the band gap offset. The effect of SAW manifests in the formation of the Floquet sub-bands (grey lines) separated by small dynamical energy gaps. To better understand such a sub-band structure, we initially assume  no SAW potential applied, (i.e., $u_0=0$) and set the modulation wavelength to the carbon-carbon bond length ($L_y=a_{CC}$). In this case we recover unperturbed SWCNT electronic structure where the conduction and valence electron sub-bands are defined by single quantum number $n=0,\pm 1, \dots$ ((Eq.~\ref{kx})).  Turning on weak SAW field results in the formation of small gaps within each dispersion curve associated with $n=0,\pm 1, \dots$ of unperturbed electronic states. This effect is illustrated in Fig.~\ref{Fig:1}a by black solid curves. In addition, increase of the SAW wavelengths from $a_{CC}$ will generate the Floquet replicas (grey curves) due to the scattering on the Bragg grating. These states are  numerated by the quantum number $s$ acquiring odd (even) integers for conduction (valence) band.

\begin{figure}[t]
  \centering
  \includegraphics[width=0.8\columnwidth,clip]{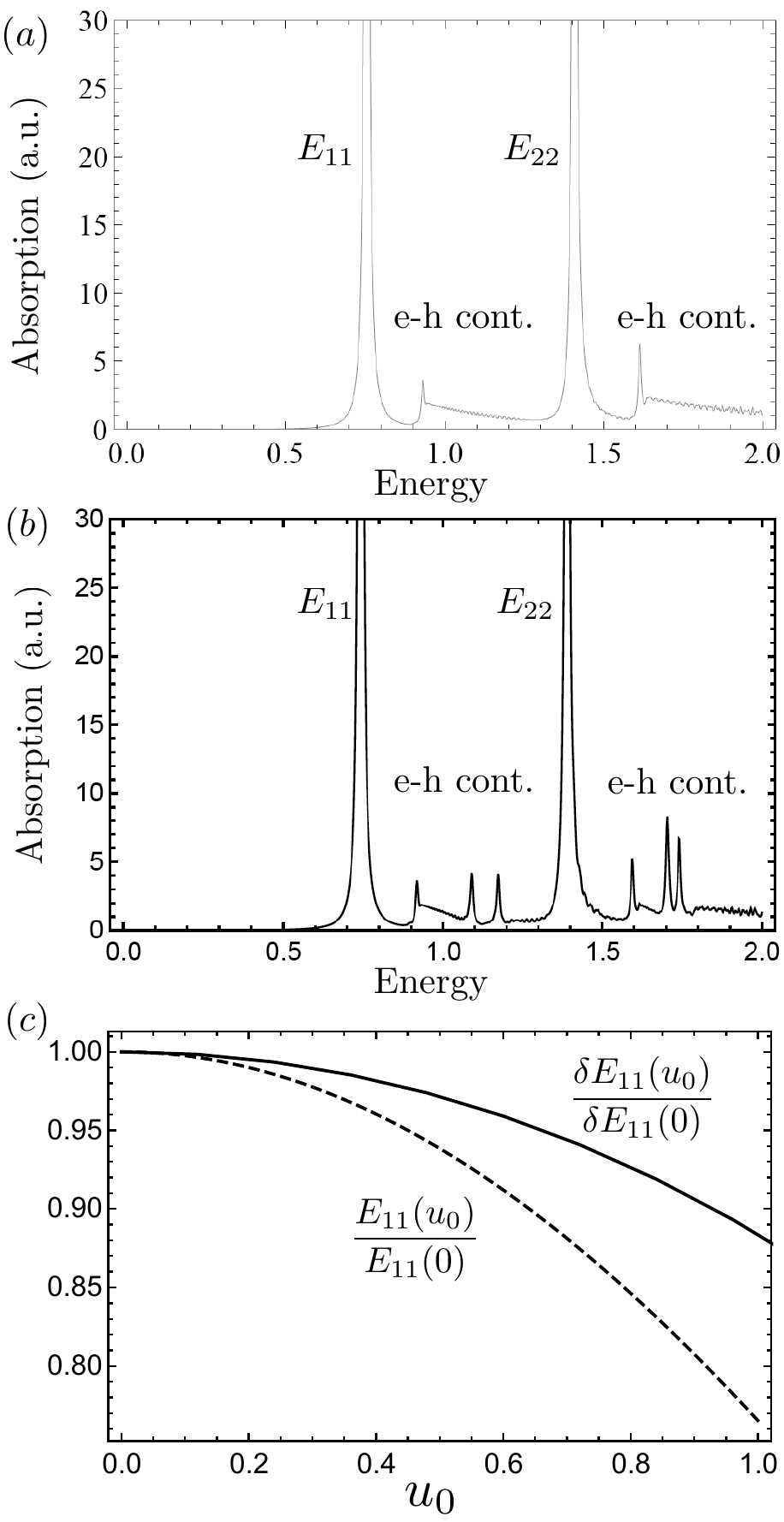}
  \caption{
Absorption spectra associated with correlated electron-hole states covering region of the lowest exciton states $E_{11}$ and $E_{22}$ and electron-hole continuum calculated (a) in the absence of SAW potential and (b) in the presence of SAW potential with the period $L_y=100L_x$ and amplitude $u_0=0.2$. Adopted energy units are the same as in Fig.~\ref{Fig:1}a. 
(c) Ratio of the exciton (binding) energy in the presence of SAW potential, $E_{11}(u_0)$ ($\delta E_{11}(u_0)$) and its absence $E_{11}(0)$ ($\delta E_{11}(0)$) vs. SAW potential amplitude. 
}
  \label{Fig:2}
\end{figure}

Optical properties of the Floquet states are determined by the transition current matrix element of non-interacting electrons  defined in Eq.~(\ref{je-mtrx}). Our numerical simulations confirm that vertical, $k_{\texttt c}=k_{\texttt v}$, optical transitions occur between the Floquet sub-bands with $s_{\texttt c}=s_{\texttt v}\pm 1$ and  $n_{\texttt c}=n_{\texttt v}$.  Among all of these states the dominant contribution comes from those which directly originate from the unperturbed states of SWCNT (solid black in panel~a of  Fig.~\ref{Fig:1}).  Below, we refer to these states as {\it optically active}  and distinguish them by their quantum number $n$. To demonstrate how the transition current matrix element depends on $k$ for optically active states, we introduce the following quantity
\begin{equation}
\label{j00}
j^\nu_{n_{\texttt c}n_{\texttt v}}(k)\sim \sum_{l_{\texttt c}l_{\texttt v}} j^\nu_{\alpha_{\texttt c}\alpha_{\texttt v}},
\end{equation}
where $k=k_{\texttt c}=k_{\texttt v}$,  $j^\nu_{\alpha_{\texttt c}\alpha_{\texttt v}}$ is given by Eq.~(\ref{je-mtrx}), and the indices $\alpha_{\texttt c}=\{s_{\texttt c},n_{\texttt c},k_{\texttt c}-l_{\texttt c}G_0\}$ and $\alpha_{\texttt v}=\{s_{\texttt v},n_{\texttt v},k_{\texttt v}-l_{\texttt v}G_0\}$ satisfy the selection rules defined above further restricted to the optically active states. Fig.~\ref{Fig:1}b shows the dispersion of the transition current defined in Eq.~(\ref{j00}) for the optically active valence and conduction band states with $n_{\texttt c}=n_{\texttt v}=0$. Sharp dips indicate the reduction in the matrix element value due to the SAW-induced dynamical gaps. Our data analysis (not shown in the plot) reveals that the current matrix element sign changes at the $k$ values associated with the dips. This indicates electron scattering effect on the  edges  of Brillouin sub-zones, i.e., $k=lG_0$, resulting in the formation of Floquet sub-bands. 

Now, we turn our attention to the correlated electron-hole states defined by the Bethe-Salpeter equation (Eq.~(\ref{X-HF})). Specifically, we solved Eqs.~(\ref{X-HF}) and (\ref{gX}) numerically using the discretization step of $0.01 G_0$ and neglecting the Coulomb exchange interaction. The eigenstates were further used in Eq.~(\ref{jX}) and convoluted with the matrix elements of the transition current due to the {\it optically active} states. This allowed us to evaluate the  absorption spectra according to  Eq.~(\ref{Spct}). Fig.~\ref{Fig:2}a presents such a spectrum of SWCNT in the absence of SAW potential. Well-know lowest $E_{11}$ and $E_{22}$ exciton states\cite{Dresselhaus:2007} are clearly seen in the plot red-shifted from the associated electron-hole continua. The difference between the exciton energy and the van Hove singularity of associated electron-hole continuum  band edge defines the exciton binding energy, $\delta E_{11}$ or $\delta E_{22}$.

\begin{figure}
  \centering
  \includegraphics[width=0.8\columnwidth]{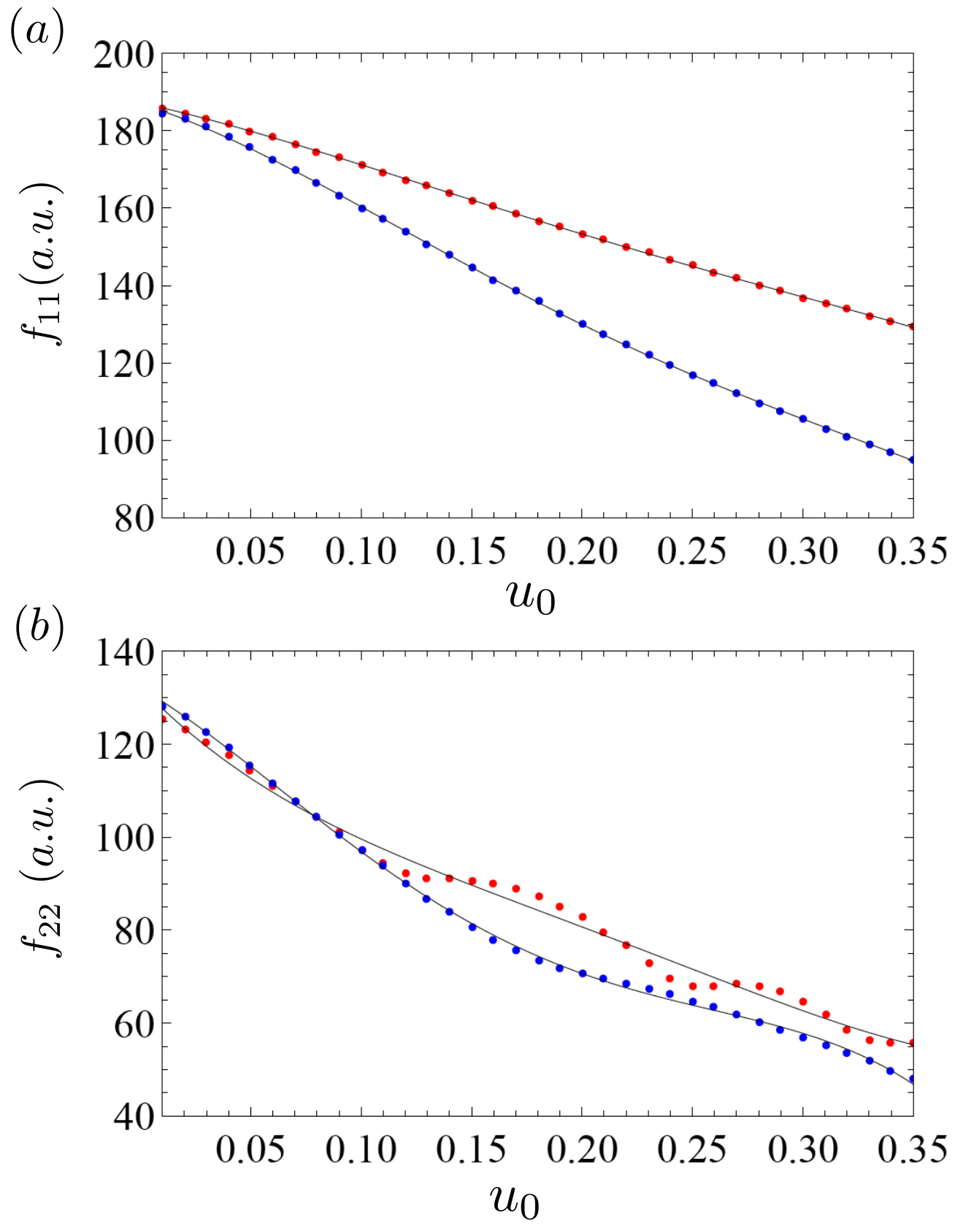}
  \caption{
Respectively, the oscillator strengths $f_{11}$ and $f_22$  of the lowest exciton states $E_{11}$ and $E_22$ plotted vs SAW amplitude, $u_0$ for different periods  
$L_y=100 L_x$ (red) and $L_y=200 L_x$ (blue).
}
  \label{Fig:3}
\end{figure}

As can be easily noticed in Fig.~\ref{Fig:2}b, turing the SAW potential on results in the appearance of addition van Hove singularities within the electron-hole continua. These singularities  mark the Floquet band-edge states appearing as a result of dynamical gap opening. More close comparison of panels a and b of Fig.~\ref{Fig:2} shows that the exciton states experience the SAW induced red shift attributed to the dynamical Stark shift  and reduction in their binding energy.  In panel c of  Fig.~\ref{Fig:2}, we plot normalized lowest exciton state energy $E_{11}(u_0)$  as a function of the SAW potential amplitude, $u_0$ (dash line). This plot clearly demonstrates the increase in the Stark shift  that according to our analysis satisfies the following simple relationship    
\begin{equation}
\label{E-stark}
E_{11}(u_0)=J_0(u_0)E_{11}(0),
\end{equation}
where $E_{11}(0)$ is the lowest exciton state energy calculated at $u_0=0$ and $J_0$ is the first kind Bessel function with $l=0$. Variation in the normalized exciton binding energy, $\delta E_{11}(u_0)$ is also shown in this plot (solid line).  It is clear, that the binding energy drops as the SAW potential increases facilitating exciton dissociation processes. Due to complex many-body effects contributing to the exciton binding energy, we cannot provide a simple analytical relationship describing the observed trend.  

Finally, we examine how SAW potential affects the exciton transition oscillator strength, $f_{11}$ and $f_{22}$ defined by Eq.~(\ref{fX}) for the lowest exciton transitions $E_{11}$ and $E_{22}$, respectively. Experimentally, the former (the later) quantity can be probed in the  photoluminescence (absorption) measurements. Fig. ~\ref{Fig:3} shows the decrease of the  oscillator strengths as the  SAW amplitude $u_0$ increases. Our analysis reveals that the oscillator strength gets transferred to the electron-hole continua, specifically to the new van Hove singularities (i.e., Floquet band-edge states) appearing in Fig.~\ref{Fig:2}b. Furthermore, the reduction in the exciton oscillator strength can be rationalized in terms of the the transition current matrix elements reduction (dips in Fig.~\ref{Fig:1}b ). The increase in the SAW potential period (compare the red and blue curves) further decreases the oscillator strengths. This effect is well pronounced for the $E_{11}$ exciton state (panel~a). For the  $E_{22}$ states (panel~b) the effect is weaker. Oscillations observed in panel~b for large  values of $u_0$ reflect numerical instabilities in solution of the Bether-Salpeter equation.\footnote{Generally, going beyond the perturbation limit in the SAW poses significant challenges. Indeed, due to small values of $G_0$ we have to maintain a fine mesh in $k-$ space in order to account for the Bragg scattering. The iterative Krylov method is applied to the matrix of $10^6\times 10^6$ size. For large SAW amplitudes  matrix in the Bether-Salpeter equation become rather dense resulting in the oscillations of $f_{22}$.} 

\section{Conclussions}
\label{Sec-Conc}

We have developed a self-consistant theoretical approach to describe optical properties of exciton states in semiconductor SWCNTs in the presence of a periodic potential that can be induced by SAW propagating along a  SWCNT. The analysis of non-interacting electronic states within the valance and conduction bands shows that the periodic potential results int the formation of Floquet sub-bands separated by dynamical energy gaps. Subsequent analysis of the correlated electron-hole states including the low energy exciton states shows that Floquet band-edge states play significant role determining optical transition in the spectral region of interest. Specifically, we predict formation of new van Hove singularities in the electron-hole continua due to the optical transitions between Floquet band edge states. These transitions acquire additional oscillator strengths at the coast of reduction in the exciton oscillator strength. The latter is a result of the electron Bragg scattering at the edges of SAW induced Brillouin sub-zones.  We also find that the SAW potential causes exciton energy red shift interpreted as a dynamical Stark effect and reduction in the exciton binding energy that can facilitate exciton state ionization and decay into the electron-hole continuum. Although general trends for the exciton oscillator strength and binding energy are in agreement with previous theoretical\cite{Perebeinos:2007} and experimental \cite{Regler:2013} studies, we clearly demonstrates the key role of the Floquet states in determining exciton optical properties that has not to be emphasized before.     

\acknowledgements O.R. acknowledges the support provided by the Fordham University startup funds. AP acknowledges the support provided by Los Alamos National Laboratory Directed Research and Development (LDRD) Funds.




\begin{thebibliography}{10}

\bibitem{Saito.Book:1998}
R. Saito, G. Dresselhaus, and M. Dresselhaus, {\em Physical Properties of
  Carbon Nanotubes} (Imperial College Press, London, 1998).

\bibitem{Dresselhaus.Book:2001}
{\em Carbon Nanotubes Synthesis, Structure, Properties, and Aplications},
  Vol.~80 of {\em Topics in Applied Physics}, edited by M. Dresselhaus, G.
  Dresselhaus, and P. Avouris (Springer, New York, 2001).

\bibitem{CastroNeto:2009}
A.~C. Neto {\it et~al.}, Rev. Mod. Phys. {\bf 81},  109  (2009).

\bibitem{Dresselhaus:2007}
M.~S. Dresselhaus, G. Dresselhaus, R. Saito, and A. Jorio, Annu. Rev. Phys.
  Chem. {\bf 58},  719  (2007).

\bibitem{Ostrovskii:1999}
I.~V. Ostrovskii, O. Korotchenkov, T. Goto, and H. Grimmeiss, Phys. Rep. {\bf
  311},  1  (1999).

\bibitem{GarciaCristobal:2004}
A. Garc{\'\i}a-Crist{\'o}bal, A. Cantarero, F. Alsina, and P. Santos, Phys.
  Rev. B {\bf 69},  205301  (2004).

\bibitem{Kinzel:2011}
J.~B. Kinzel {\it et~al.}, Nano Lett. {\bf 11},  1512  (2011).

\bibitem{Rocke:1997}
C. Rocke {\it et~al.}, Phys. Rev. Lett. {\bf 78},  4099  (1997).

\bibitem{Santos:1998}
P. Santos, M. Ramsteiner, and F. Jungnickel, Appl. Phys. Lett. {\bf 72},  2099
  (1998).

\bibitem{Volk:2010}
S. V\"olk {\it et~al.}, Nano Lett. {\bf 10},  3399  (2010).

\bibitem{Volk2011}
S. V\"olk {\it et~al.}, Appl. Phys. Lett. {\bf 98},  023109  (2011).

\bibitem{Hernandez:2011}
A. Hern{\'a}ndez-M{\'\i}nguez {\it et~al.}, Nano Lett. {\bf 12},  252  (2011).

\bibitem{Gell:2008}
J. Gell {\it et~al.}, App. Phys. Lett. {\bf 93},  081115  (2008).

\bibitem{Fuhrmann:2011}
D.~A. Fuhrmann {\it et~al.}, Nature Photonics {\bf 5},  605  (2011).

\bibitem{DeLima:2005}
M.~M. de~Lima~Jr and P.~V. Santos, Rep. Progr. Phys. {\bf 68},  1639  (2005).

\bibitem{Talyanskii:2001}
V. Talyanskii, D. Novikov, B. Simons, and L. Levitov, Phys. Rev. Lett. {\bf
  87},  276802  (2001).

\bibitem{Novikov:2005}
D.~S. Novikov, Phys. Rev. B {\bf 72},    (2005).

\bibitem{Thouless:1983}
D. Thouless, Phys. Rev. B {\bf 27},  6083  (1983).

\bibitem{Buitelaar:2008}
M. Buitelaar {\it et~al.}, Phys. Rev. Lett. {\bf 101},  126803  (2008).

\bibitem{Leek:2005}
P. Leek {\it et~al.}, Phys. Rev. Lett. {\bf 95},  256802  (2005).

\bibitem{Wurstle:2007}
C. Wurstle, J. Ebbecke, M. Regler, and A. Wixforth, New J. Phys. {\bf 9},  73
  (2007).

\bibitem{Regler:2013}
M. Regler {\it et~al.}, Chem. Phys. {\bf 413},  39  (2013).

\bibitem{Perebeinos:2007}
V. Perebeinos and P. Avouris, Nano Lett. {\bf 7},  609  (2007).

\bibitem{Ando:1997}
T. Ando, J. Phys. Soc. Jpn. {\bf 66},  1066  (1997).

\bibitem{Ando:2005}
T. Ando, J. Phys. Soc. Jpn. {\bf 74},  777  (2005).

\bibitem{Gu:2011}
Z. Gu, H. Fertig, D.~P. Arovas, and A. Auerbach, Phys. Rev. Lett. {\bf 107},
  216601  (2011).

\bibitem{Perez:2014}
P. Perez-Piskunow, G. Usaj, C. Balseiro, and L.~F. Torres, Phys. Rev. B {\bf
  89},  121401  (2014).

\bibitem{Kushwaha:2001}
M.~S. Kushwaha, Surf. Sci. Rep. {\bf 41},  1  (2001).

\bibitem{Jiang:2007}
J. Jiang {\it et~al.}, Phys. Rev. B {\bf 75},  035407  (2007).

\bibitem{Spataru:2004}
C.~D. Spataru, S. Ismail-Beigi, L.~X. Benedict, and S.~G. Louie, Phys. Rev.
  Lett. {\bf 92},  077402  (2004).

\end{thebibliography}

\end{document}